\documentclass[dvipdfmx,aps,prl,reprint,groupedaddress]{revtex4-1}

\usepackage{graphicx,color}

\def\U#1{{%
\def\O{\mbox{O}}
\def\u{\mbox{u}}
\mathcode`\u=\mu
\mathcode`\O=\Omega
\mathrm{#1}}}
\def\ii{{\mathrm{i}}}
\def\ee{{\mathrm{e}}}
\def\dd{{\mathrm{d}}}

\begin{document}


\title{Integer Multiplier for Orbital Angular Momentum of Light \\using Circular-Sector Transformation}


\author{Satoru Takashima}
\author{Hirokazu Kobayashi}
\email{kobayashi.hirokazu@kochi-tech.ac.jp}
\author{Katsushi Iwashita}
\affiliation{Graduate School of Engineering, Kochi University of Technology, 185 Tosayamada-cho, Kochi 782-8502 Japan}


\date{\today}

\begin{abstract}
This paper describes an integer multiplier for the orbital angular momentum (OAM) of light through the parallel implementation of multiple circular-sector transformations, whereby the cross-sectional circular shape of the OAM mode is geometrically transformed to the circular-sector shape.
Experiments show that the conversion accuracy of both OAM doubler and tripler formulations is significantly better than that of  the previous method.
This is because the proposed method uses a simple implementation with a single spatial light modulator.
The proposed method has strong potential for the spatial mode manipulation of OAM and other useful spatial modes.
\end{abstract}


\maketitle

It is more than 20 years since Allen \textit{et al.} recognized that light waves with an azimuthal phase term $\exp(\ii l\theta)$ are associated with photons that have a quantized intrinsic orbital angular momentum (OAM) $l\hbar$, where $\theta$ is the azimuthal angle on the beam cross-section and the integer $l$ is the topological charge\cite{allen1992orbital}.
The unlimited range of the topological charge $l$ brings a new degree of freedom with an unbounded state space for light waves, and hence the mode of the OAM has become a useful tool in numerous applications\cite{rubinsztein2016roadmap}, such as spatial mode multiplication\cite{gibson2004free,wang2012terabit}, microscopy\cite{hell1994breaking}, optical tweezers\cite{simpson1996optical}, high-dimensional entanglement\cite{mair2001entanglement,fickler2012quantum}, and quantum cryptography\cite{groblacher2006experimental,mafu2013higher}.

The creation or manipulation of OAM modes can be accomplished using various optical elements, e.g., spiral phase plates\cite{beijersbergen1994helical,harm2015adjustable}, computer-generated holograms\cite{heckenberg1992generation,ando2009mode}, $q$-plates\cite{marrucci2006optical,marrucci2013q}, conical mirrors\cite{mansuripur2011spin,kobayashi2012helical}, and metamaterials or metasurfaces\cite{yu2011light,zhao2013metamaterials,karimi2014generating}.
However, these conventional methods can only perform shift operations (additive or subtractive operations) on the OAM mode.
In addition to shifting the OAM mode, it would be extremely useful to be able to multiply the OAM state for some applications, such as the multiplicative creation of higher-order OAM modes, optical switching/routing operations\cite{willner2016orbital}, and optical information processing\cite{garcia2008quantum}.
Although an OAM multiplier combined with frequency up-conversion has been achieved using nonlinear second harmonic generation\cite{dholakia1996second,bloch2012twisting,gariepy2014creating}, this is unlikely to be sufficiently efficient for many practical applications.

Recently, another sophisticated implementation of OAM multiplication was reported using log-polar geometric transformations\cite{potovcek2015quantum,zhao2016invited}, which had previously been exploited for OAM mode sorting\cite{berkhout2010efficient,o2012near,wen2018spiral}.
In the log-polar OAM multiplier, the annular shape of the OAM modes is unwrapped to $N$ copies of the rectangular shape with an $N$-fold linear phase, followed by rewrapping to the annular shape.
This method has the excellent property that, in principle, lossless and reversible conversion is possible.
However, there is also a fundamental problem, as the accuracy of the OAM multiplication decreases because of experimental difficulties in implementing multiple log-polar geometric transformations and the loss of the periodic boundary condition along the azimuthal angle of the OAM modes.

In this letter, we consider an essentially different approach that avoids the limitations of the log-polar OAM multiplier.
We propose and experimentally demonstrate that \textit{circular-sector transformation}, i.e., mapping the circular to the circular-sector shape, can be exploited as an OAM multiplier.
The proposed OAM multiplier is based on the parallel implementation of multiple circular-sector transformations using the double-phase hologram technique\cite{hsueh1978computer,mendoza2014encoding}.
Our method enables highly accurate and highly efficient OAM multiplication with a simple setup and single-step geometric transformation, without loss of the azimuthal periodic boundary condition. The theoretical predictions are verified by numerical simulations and experiments.

A typical optical system for the geometric transformation or coordinate mapping, proposed by Bryngdahl\cite{bryngdahl1974geometrical,bryngdahl1974optical}, is the $2f$ configuration with two phase masks. The first is the transforming phase $\varphi(x,y)$, which implements the geometric transformation placed at the front focal plane $(x,y)$ of the Fourier-transforming lens, and the second one is the correction phase $\Psi(u,v)$, which compensates the undesired phase in the transformed beam at the Fourier plane $(u,v)$.
In approximating the Fourier transform integral using the stationary-phase method, the point $(x,y)$ is mapped onto the point $(u,v)$ given by
\begin{equation}
(x,y)\mapsto (u,v)=\frac{f}{k}\left(\varphi_x,\varphi_y\right),
\label{eq:u,v,x,y}
\end{equation}
where $f$ is the focal length of the Fourier-transforming lens, $k$ is the wave number, and the subscripts $x$ and $y$ denote partial differentiation with respect to $x$ and $y$, respectively.
For geometric transformation applications, a mapping $(x,y)\mapsto (u(x,y),v(x,y))$ is given, and then the transforming phase $\varphi(x,y)$ is determined by solving the partial differential equations in Eq.~(\ref{eq:u,v,x,y}) when the continuity condition $u_y=v_x$ is satisfied~\cite{stuff1990coordinate}.

\begin{figure}[t]
\includegraphics[clip, keepaspectratio, width=8.6cm]{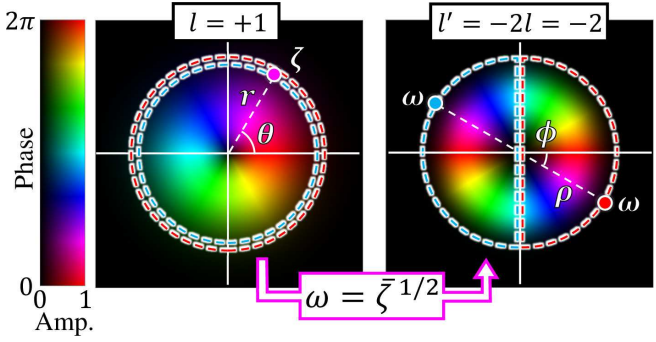}
\caption{\label{fig:2Multiplication}Principle of integer multiplication of OAM using circular-sector transformations with $N=2$ (OAM doubler). The input plane $\zeta=r\ee^{\ii\theta}$ is mapped onto the output plane $\omega=\rho\ee^{\ii\phi}$ associated with the complex fractional power function $\omega=g(\bar{\zeta})=\bar{\zeta}^\frac{1}{2}$. The red and blue dotted circular shapes are mapped onto the semicircular shapes at opposite azimuthal positions by two different circular-sector transformations.}
\end{figure}

If the complex form of the transformation $(x,y)\mapsto (u,v)$ is an anti-analytical function, i.e., $\omega=g(\bar{\zeta})$, where $\omega=u+\ii v$ and $\bar{\zeta}=x-\ii y$, then the continuity condition is satisfied by virtue of the Cauchy--Riemann differential equation for $g(\bar{\zeta})$\cite{cederquist1984computer}.
In the complex form, the solution of Eq.~(\ref{eq:u,v,x,y}) can be expressed in simple form\cite{bereznyi1990synthesized} as
\begin{equation}
\varphi(x,y)=\frac{k}{f}\mathrm{Re}\left[\int g\left(\bar{\zeta}\right)\dd\bar{\zeta}\right].
\label{eq:varphi}
\end{equation}

We now consider the circular-sector transformation for implementing the OAM multiplier.
The optical complex amplitude $E_l(r,\theta)$ of OAM modes with topological charge $l$ can be expressed in polar coordinates $(r,\theta)$ as 
$E_l(r,\theta)\propto E_0(r)\left(r\ee^{\ii\theta}\right)^l\propto \zeta^l$, where the complex variable $\zeta=x+\ii y=r\ee^{\ii\theta}$ and $E_0(r)$ is the complex amplitude of the fundamental Gaussian mode.
The circular-sector transformation exploits a fractional power function as an anti-analytic complex function $g(\bar{\zeta})$ expressed as
\begin{equation}
\omega=g(\bar{\zeta})=\alpha\bar{\zeta}^\frac{1}{N},
\label{eq:omega}
\end{equation}
where the integer $N$ corresponds to the factor of the OAM multiplier and $\alpha$ is a real-valued scaling constant.
The geometric transformation of the OAM modes by Eq.~(\ref{eq:omega}) can be formulated as 
$\zeta^l\mapsto\left(\bar{\omega}/\alpha\right)^{Nl}\propto\left(\rho\ee^{-\ii\phi}\right)^{Nl}$ on the output polar coordinate $(\rho,\phi)$. Thus, the input OAM of $l$ is multiplied by $-N$.
Note that the fundamental Gaussian amplitude $E_0(r)$ is also geometrically converted and becomes a super-Gaussian function.
However, using the appropriate spatial low-pass filter, the super-Gaussian function can be approximated as a fundamental Gaussian function.

The complex fractional power function in Eq.~(\ref{eq:omega}) has a branch point of order $N$ at the origin and is an $N$-valued function.
Thus, the input polar coordinates $(r,\theta)$ are mapped onto the $N$ points on the output polar coordinates, 
$(\rho,\phi)=\left(\alpha r^\frac{1}{N},-\frac{\theta+2n\pi}{N}\right)$, where the integer $n=0,\cdots,N-1$.
As shown in Fig.~\ref{fig:2Multiplication}, each coordinate mapping with a particular value of $n$ converts the circular shape of the input OAM mode to the circular-sector shape at a different azimuthal position.
If all $N$ circular-sector transformations occur simultaneously, the input OAM is multiplied by $-N$ in the output.
From Eqs.~(\ref{eq:varphi}) and (\ref{eq:omega}), the transforming phase $\varphi_n(r,\theta)$ for the $n$-th circular sector transformation can be calculated as
\begin{equation}
\varphi_n(r,\theta)=\frac{\alpha k}{f}\frac{Nr^{1+\frac{1}{N}}}{N+1}\cos\left[\frac{(N+1)\theta+2n\pi}{N}\right].
\label{eq:varphi phase}
\end{equation}

The required complex amplitude modulation for the parallel implementation of the $N$ circular-sector transformations is given by
\begin{equation}
A(r,\theta)\ee^{\ii\varphi(r,\theta)}=\frac{1}{N}\sum_{n=0}^{N-1}\ee^{\ii\varphi_n(r,\theta)},
\label{eq:sigma phase distribution}
\end{equation}
where $A(r,\theta)$ and $\varphi(r,\theta)$ represent the normalized amplitude and phase distribution, respectively.
For a simple and efficient implementation of Eq.~(\ref{eq:sigma phase distribution}) with single-phase-only spatial light modulator (SLM), we exploit the double-phase hologram technique\cite{hsueh1978computer,mendoza2014encoding}, whereby the required phase $\Phi(r,\theta)$ is formulated as
\begin{equation}
\Phi(r,\theta)=\varphi(r,\theta)+\Pi_\pm\cos^{-1}A(r,\theta),
\label{eq:phase distribution}
\end{equation}
where $\Pi_\pm=(-1)^{n+m}$ with the $x-$ and $y-$directional pixel numbers $n$ and $m$ of the SLM is a spatially periodic function that returns values of $+1$ or $-1$, like a two-dimensional binary grating (checker-board pattern).
The complex amplitude modulation in Eq.~(\ref{eq:sigma phase distribution}) can be accomplished by applying a spatial low-pass filter to extract the zeroth-order diffraction component from the product of the input OAM modes and the phase mask $\Phi(r,\theta)$.

Finally, the undesired phase factor of the transformed complex amplitude is compensated by the correction phase $\Psi(\rho,\phi)$ calculated under the stationary-phase method\cite{bryngdahl1974geometrical,stuff1990coordinate}.
As the circular-sector shapes do not overlap, the correction phase can be formulated as the following single continuous function:
\begin{equation}
\Psi(\rho,\phi)=-\frac{k}{f}\left(\frac{\rho}{\left|\alpha\right|}\right)^N\frac{\rho\cos\left[(N+1)\phi\right]}{N+1}.
\label{eq:psi phase}
\end{equation}

\begin{figure}[t]
\includegraphics[clip, keepaspectratio, width=8.65cm]{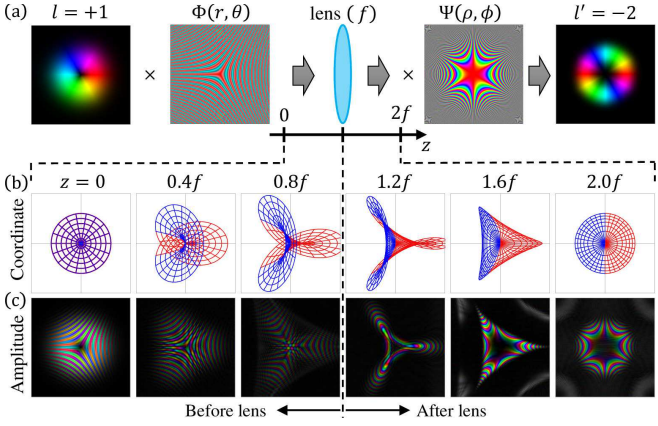}
\caption{\label{fig:Conversion process}Numerical simulation of conversion process of OAM multiplication with $N=2$ (OAM doubler). (a) Two required phase masks, $\Psi$ and $\Phi$, for the OAM doubler and output complex amplitude calculated by fast Fourier transform. (b) Ideal conversion process of the two polar coordinates (red and blue) by different circular sector transformations. (c) Conversion process of complex amplitude of the input OAM mode calculated by Fresnel diffraction integral.}
\end{figure}
Figure~\ref{fig:Conversion process} shows the results of numerical simulations for the OAM doubler (see SM\cite{suppl} for details of the calculations).
Figure~\ref{fig:Conversion process}(a) presents the simulation result obtained by calculating the fast Fourier transform of the input OAM modes with $l=+1$ multiplied by the transforming phase.
After compensating the undesired phase by $\Psi(\rho,\phi)$, the input OAM is doubled as $l^\prime=-2l$.
Figure~\ref{fig:Conversion process}(b) illustrates the ideal conversion process of the polar coordinate system during propagation inside the $2f$ system. 
The red and blue polar coordinates are geometrically transformed by different circular-sector transformations with $n=0$ and $n=1$, resulting in a semicircular shape at opposite azimuthal positions.
Using the double-phase hologram $\Phi(r,\theta)$ in Eq.~(\ref{eq:phase distribution}), these two coordinate transformations occur simultaneously, and thus, the two coordinates do not break the periodic boundary condition along the azimuthal angle.
From the viewpoint of complex function theory, the OAM multiplier geometrically unwraps the Riemann surface for the complex fractional power function to a single leaf.
Figure~\ref{fig:Conversion process}(c) shows the conversion process of the complex amplitude obtained by calculating the Fresnel diffraction in the angular spectrum domain.
After propagation over a distance of $z=2f$, the optical amplitude distribution has an annular shape, but is accompanied by an undesired phase, which is compensated by the correction phase $\Psi(\rho,\phi)$.

\begin{figure}[t]
\includegraphics[clip, keepaspectratio, width=8.6cm]{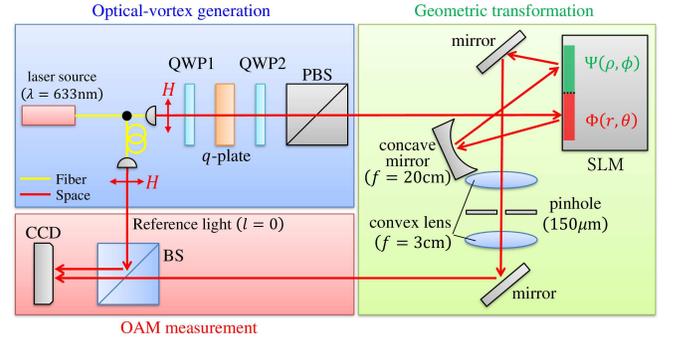}
\caption{\label{fig:overview}Schematic of the experimental setup. BS, PBS, and QWP denote beam splitter, polarization BS, and quarter waveplate, respectively.}
\end{figure}

The experimental setup is shown in Fig.~\ref{fig:overview}.
A light wave irradiated from a single-mode fiber pigtailed laser diode with a wavelength of $635\,\U{nm}$ is split into two paths by the fiber coupler, one for preparing the OAM modes and the other for the reference beam.
The OAM modes ($-2\leq l\leq 2$) and the balanced superpositions of the positive and negative OAM modes are generated from the $2.5$-$\U{mm}$-radius collimated Gaussian beam using a $q$-plate device with a topological charge of $q=\frac{1}{2}$ or $1$.

For the circular sector transformations, the two required phase masks $\Phi(r,\theta)$ and $\Psi(\rho,\phi)$, with the parameter $\alpha=0.025\,\U{m}^{1/2}$ for the OAM doubler and $\alpha=0.0092\,\U{m}^{2/3}$ for the OAM tripler, are prepared on the halves of the single SLM.
The input OAM mode subjected to the transforming phase $\Phi(r,\theta)$ is Fourier transformed in the $2f$ configuration via reflection by the concave mirror with focal length $f=200\,\U{mm}$, followed by compensation with $\Psi(\rho,\phi)$.
(In the actual experiments, a linear blazed grating phase was included in the transforming phase.) Finally, the spatial filter composed of two lenses and a pinhole extracts the first-order diffraction component, thus achieving complex amplitude modulation in Eq.~(\ref{eq:varphi}) and reducing the undesired diffraction noise.

In the OAM measurement stage, the intensity distributions of the transformed beam and its interferograms with the reference beam are observed by a CCD camera.
To verify whether the desired OAM multiplication can be implemented, the complex amplitude distribution is reconstructed from the interferogram using the angular spectrum method\cite{takeda1982fourier}.

\begin{figure}[t]
\includegraphics[clip, keepaspectratio, width=8.8cm]{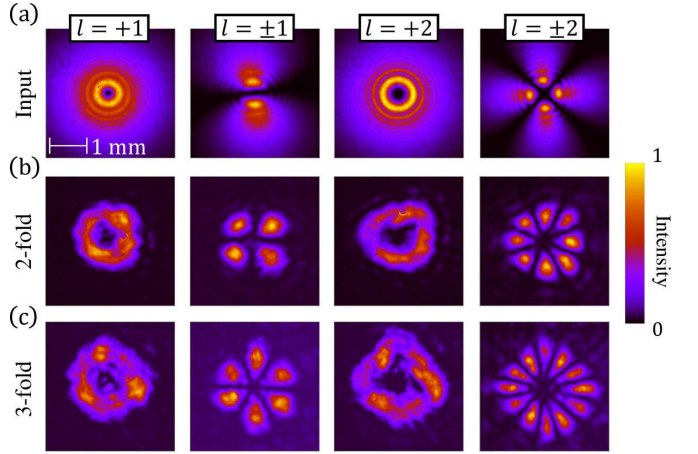}
\caption{\label{fig:Experimental results of 2-fold and 3-fold}Intensity distribution of (a) input OAM modes, (b) 2-fold, and (c) 3-fold OAM modes obtained experimentally by geometric transformation.}
\end{figure}
\begin{figure}[t]
\includegraphics[clip, keepaspectratio, width=8.6cm]{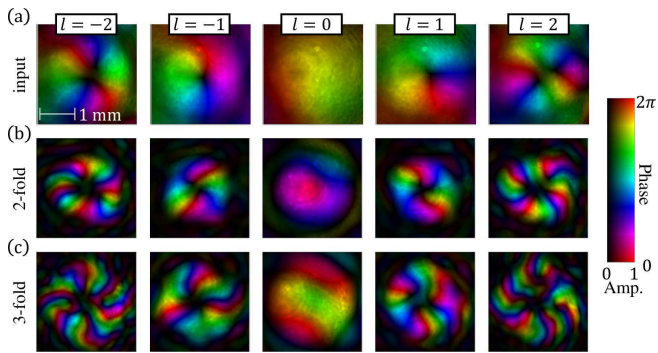}
\caption{\label{fig:Complex amplitude of the multiplied optical vortex}Complex amplitude distribution obtained experimentally after 2-fold and 3-fold OAM multiplication. Complex amplitude distribution of (a) input, (b) 2-fold, and (c) 3-fold OAM beams.}
\end{figure}
Figure~\ref{fig:Experimental results of 2-fold and 3-fold} shows the experimentally obtained intensity distributions of the input, doubled, and tripled OAM modes.
The input OAM modes with $l=\pm 1, \pm 2$ represent balanced superpositions of $+l$ and $-l$, with $l=1,2$, in which the intensity distribution displays $2|l|$ intensity maxima or ``petals'' along the azimuthal angle.
As shown in the second and fourth columns in Fig.~\ref{fig:Experimental results of 2-fold and 3-fold}, we can successfully obtain a geometric transformation resulting in $2N|l|$ petals while maintaining highly accurate rotational symmetry and high visibility.
For the single OAM mode with $l=+1,+2$, the annular intensity distribution is as shown in the first and third columns of Fig.~\ref{fig:Experimental results of 2-fold and 3-fold}.

Figure~\ref{fig:Complex amplitude of the multiplied optical vortex} shows the complex amplitude distributions extracted from the interferogram of the input OAM modes $(-2\leq l \leq 2)$ and of the transformed modes given by the OAM doubler and tripler.
As expected, the azimuthal phase variation in the input OAM modes becomes doubled or tripled and has a negative sign.
In the case of the fundamental Gaussian input $(l=0)$, its phase structure remains almost unchanged.

\begin{figure}[t]
\includegraphics[clip, keepaspectratio, width=8cm]{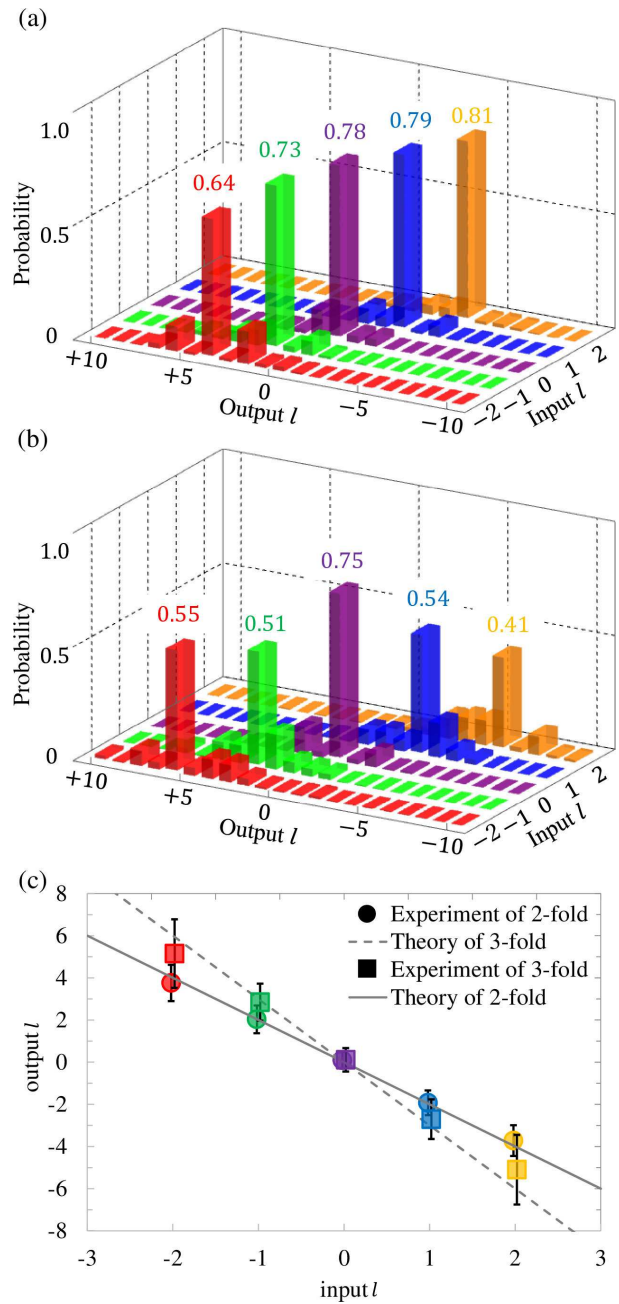}
\caption{\label{fig:mode analysis}Experimental results of OAM mode analysis. Probability distribution of OAM with (a) 2-fold and (b) 3-fold OAM multiplication results. (c) Average of the OAM and its standard deviation. Error bars show standard deviations of the OAM spectra in (a) and (b).}
\end{figure}
Finally, we confirm the accuracy of the OAM multiplier by calculating the OAM spectrum from the optical complex amplitude in Fig.~\ref{fig:Complex amplitude of the multiplied optical vortex} (see SM\cite{suppl} for details of the calculations).
Figure~\ref{fig:mode analysis} shows the output OAM spectra of the OAM doubler and tripler, in which the vertical axis shows the intensity ratio of a particular OAM mode to all OAM modes within the topological charge $-10\leq l\leq 10$.
The ratio of the desired OAM mode reaches almost $70\%$ for the OAM doubler and almost $50\%$ for the OAM tripler, and is more than four times the ratio of undesired other OAM modes, even for the OAM tripler.
Figure~\ref{fig:mode analysis}(c) shows the averaged OAM (filled circles and filled squares) and its standard deviation (error bar) calculated from the OAM spectra in Figs.~\ref{fig:mode analysis}(a) and (b).
There is almost no overlap between each output OAM mode within the range of one standard deviation.

Using our method, the conversion accuracy of the OAM multiplier is significantly improved over that of the log-polar OAM multiplier.
This is obvious from a comparison of the experimental results in Fig.~\ref{fig:mode analysis} with those for the log-polar OAM multiplier\cite{potovcek2015quantum,zhao2016invited}.
The improvement is the result of maintaining the periodic boundary condition under the geometric transformation and simplifying the experimental setup with a single SLM.
The only drawback of the proposed method is that it suffers an inevitable intensity loss because of the complex amplitude modulation in Eq.~(\ref{eq:sigma phase distribution}), meaning that the conversion efficiency cannot exceed $50\%$ when the maximum transmittance is limited to unity for passive phase elements.
We believe this drawback can be overcome by using multiple passive phase elements or some other sophisticated implementation of the complex amplitude modulation.

In summary, we have proposed a method for OAM multiplication through the parallel implementation of multiple circular-sector transformations.
First, we introduced a general solution for the geometric transformation in complex form, and then derived the circular-sector transformations by exploiting a complex fractional power function as a coordinate mapping. The theoretical predictions have been experimentally demonstrated, verifying the significant improvement in conversion accuracy for the OAM doubler and tripler when compared with the previous method. 
We expect the proposed method to be applicable to the highly accurate fractional multiplication and division of the OAM. 
Moreover, our scheme can be generalized to other sophisticated geometric transformations induced by multi-valued complex functions. 
It offers great potential for the spatial mode manipulation of the OAM and other useful spatial modes. 

The authors aknowledge G. Bateson for fruitful discussion. We would like to thank Editage for English language editing.
This work was supported by the Matsuo Foundation, Research Foundation for Opto-Science and Technology, and JSPS KAKENHI (Grant Number 18K14151). 

\providecommand{\noopsort}[1]{}\providecommand{\singleletter}[1]{#1}%

\end{document}